\documentclass[oldversion]{aa}
\usepackage{graphicx}
\begin{document}
\title{Outburst evolution, historic light curve and a flash-ionized
       nebula around the WZ~Sge-type~object PNV J03093063+2638031}
   \author{U.~Munari\inst{1},
           R.~Jurdana-{\v S}epi{\'c}\inst{2},
           P.~Ochner\inst{1}, and
           G.~Cherini\inst{3}}             

   \offprints{ulisse.munari@oapd.inaf.it}

  \institute{INAF Osservatorio Astronomico di Padova, 36012 Asiago (VI), Italy
             \and
             Physics Department, University of Rijeka, Radmile Matejcic 2, 51000, Rijeka, Croatia 
             \and
             ANS Collaboration, c/o Osservatorio Astronomico, via dell'Osservatorio 8, 36012 Asiago (VI), Italy}

   \date{Received YYY ZZ, XXXX; accepted YYY ZZ, XXXX}

     \abstract{We have monitored the 2014 superoutburst of the
WZ~Sge-type~transient PNV J03093063+2638031 for more than four months, from
$V$=11.0 maximum brightness down to $V$=18.4 mag, close to quiescence
value, by obtaining $B$$V$$R_{\rm C}$$I_{\rm C}$ photometry and low
resolution fluxed spectroscopy.  The evolution was normal and no late-time
`echo' outbursts were observed.  The absolute integrated flux of emission
lines kept declining along the superoutburst, and their increasing contrast
with the underlying continuum was simply the result of the faster decline of
the continuum compared to the emission lines.  Inspection of historical
Harvard plates covering the 1899-1981 period did not reveal previous
outbursts, neither `normal' nor 'super'.  We discovered an extended emission
nebula (radius $\sim$1 arcmin) around PNV J03093063+2638031, that became
visible for a few months as the result of photo-ionization from the
superoutburst of the central star.  It is not present on Palomar I and II
sky survey images and it quickly disappeared when the outburst was over. 
From the rate at wich the inization front swept through the nebula, we
derive a distance of $\sim$120 pc to the system.  The nebula is density
bounded with an outer radius of 0.03 pc, and the absolute magnitude of the
central star in quiescence is M$_V$$\sim$14.2 mag.  The electron density in
the nebula is estimated to be 10$^5$ cm$^{-3}$ from the observed
recombination time scale.  Given the considerable substructures seen across
the nebula, a low filling factor is inferred.  Similar nebulae have not been
reported for other WZ~Sge objects and the challenges posed to 
models are considered.

               }

   \authorrunning{U. Munari et al.}
   \titlerunning{A flash-ionized nebula around PNV J03093063+2638031}

   \maketitle

\section{Introduction}

According to the reports appeared on TOCP (Transient Object Confirmation
Page Reports) web page of the IAU Central Bureau for Astronomical 
Telegrams\footnote{http://www.cbat.eps.harvard.edu/unconf/tocp.xml}, PNV
J03093063+2638031 was discovered on 2014 Oct 29.630 UT (JD=2456960.130) as a
$\sim$11.0 mag optical transient by S.  Ueda observing from Japan with a
25cm reflector.  The astrometric position was later refined by S.  Kiyota to
have end figures 29$^s$.86 and 04$^{\prime\prime}$.49 in RA and Dec
respectively, 29$^s$.77 and 04$^{\prime\prime}$.3 by T.  Noguchi, and
29$^s$.77 and 04$^{\prime\prime}$.6 by M.  Caimmi (cf.  TOCP).  We take the
straight average of these three determinations as the best available J2000
astrometric position of the transient: $\alpha = 03^h 09^m 29^s.80 ~~(\pm
0^s.03)$ $\delta = +26^\circ 38^\prime 4^{\prime\prime}.5
~~(\pm0^{\prime\prime}.1)$ which is 12.5 arcsec away from the originally
reported position and suggests OT J030929.8+263804 as a possibly more
appropriate name for the transient.

P. Berardi (cf. TOCP) on Oct 30.8 UT and Santangelo and Gambogi (2014) also
on Oct 30.8 UT obtained low resolution spectroscopic observations of PNV
J03093063+2638031 that showed a blue continuum, narrow H-alpha line profile
and broad absorptions for the other members of Balmer lines, which suggested
the object to be a cataclysmic variable (CV) in outburst.  Not much else has
been published on this transient.  VSNET
alert-news\footnote{http://www.kusastro.kyoto-u.ac.jp/vsnet/} reported on
the detection of early superhumps in time resolved photometry of PNV
J03093063+2638031, with an amplitude of 0.025 mag and a period of 0.05613
days (= 80$^m$ 49$^s$) on Oct 31/Nov 1.  The superhumps remained visible
during following weeks.

The transient was soon classified as a new WZ Sge-type object on these
VSNET reports.  Such a classification is sound, given the presence of {\em
early} superhumps, the large amplitude of the recorded outburst and its
rarity (no previous outbursts recorded).  WZ Sge stars are an extreme case
of SU UMa-type of CVs, with intervals between superoutbursts lasting
decades, while normal outbursts are few and far in between (Hellier 2001). 
WZ Sge itself undergoes a superoutburst every $\sim$30 years (events
recorded in 1913, 1946, 1978 and 2001; Kuulker et al.  2011) and has never
been seen to undergo a normal outburst in between.  At the opposite end of
SU UMa-type of CVs are the ER UMa stars, which spend typically a third to
half of their time in superoutbursts, going into one every 1-2 months. 
Outside superoutbursts, the ER UMa stars experience a rapid series of normal
outbursts, showing one every few days.

In this paper we report about ($a$) the photometric and spectroscopic
evolution of PNV J03093063+2638031 (hereafter 'PNV' for short) during the
2014 superoutburst from our $B$$V$$R_{\rm C}$$I_{\rm C}$ photometry and
fluxed spectroscopy, ($b$) the results of our inspection of the Harvard
plate stack to reconstruct the past photometric history of the object, and
($c$) the discovery of a spatially resolved nebula around PNV, flash-ionized
by the 2014 superoutburst.

\section{Observations}

  \begin{table}[!Ht]
     \centering
     \caption{Our $B$$V$$R_{\rm C}$$I_{\rm C}$ photometry of PNV
              J03093063 +2638031 during the 2014 superoutburst. The last
              column identifies the telescopes: $a$= ANS 011, $b$= Asiago
              67/92cm Schmidt, $c$=Asiago 1.82m + AFOSC.}
     \includegraphics[width=8.8cm]{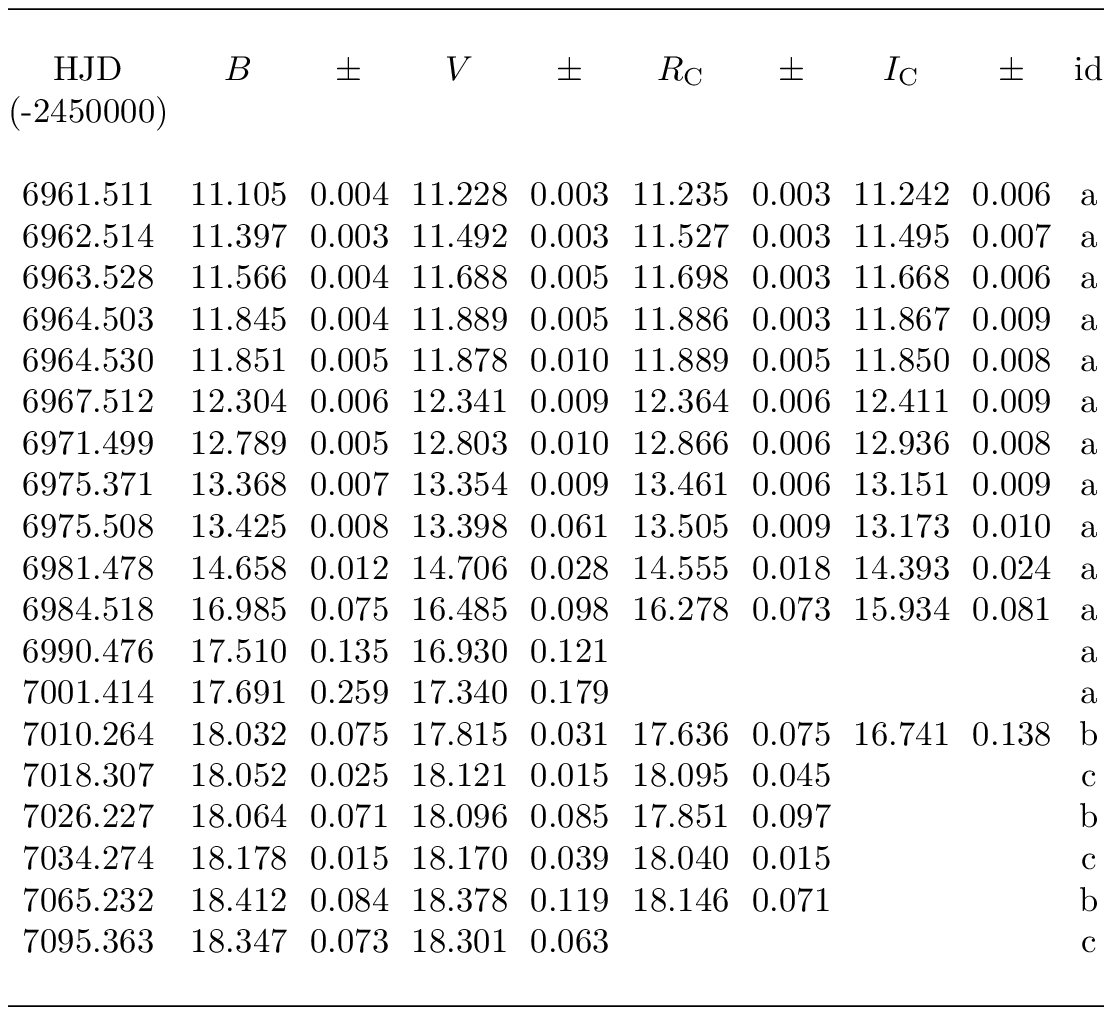}
     \label{tab1}
  \end{table}

  \begin{table}[!Ht]
     \centering
     \caption{Logbook of our spectroscopic observations of PNV
     J03093063+2638031 during the 2014 superoutburst. All spectra
     have been obtained with the Asiago 1.22m + B\&C telescope, at a
     dispersion of 2.31 \AA/pix and covering the range 3300-7950 \AA.
     $\Delta t$ is counted from optical maximum.}
     \includegraphics[width=7.6cm]{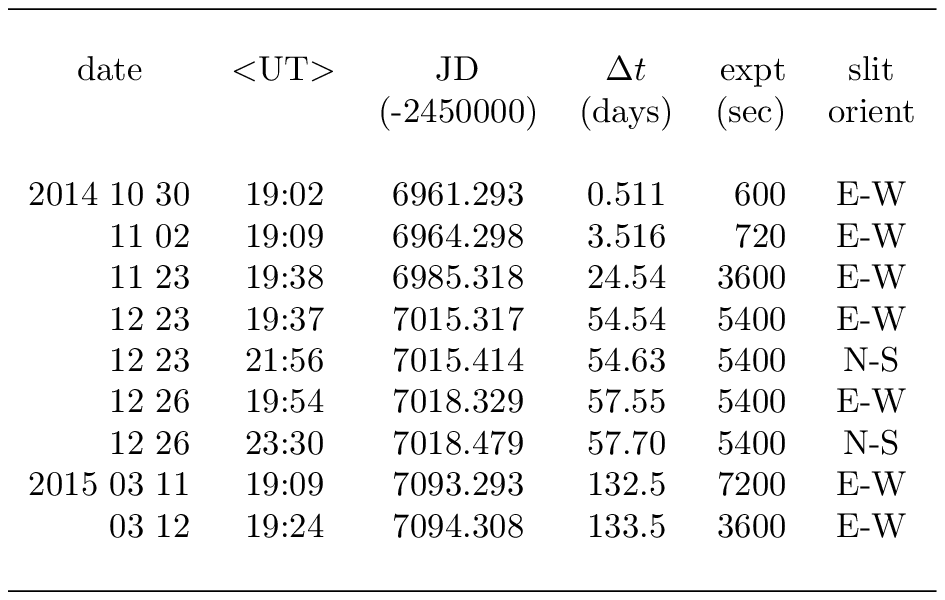}
     \label{tab2}
  \end{table}

  \begin{table}[!Ht]
     \centering
     \caption{Cross-identification of PNV J03093063+2638031 with other
     catalogs and the respective astrometric positions.}
     \includegraphics[width=8.8cm]{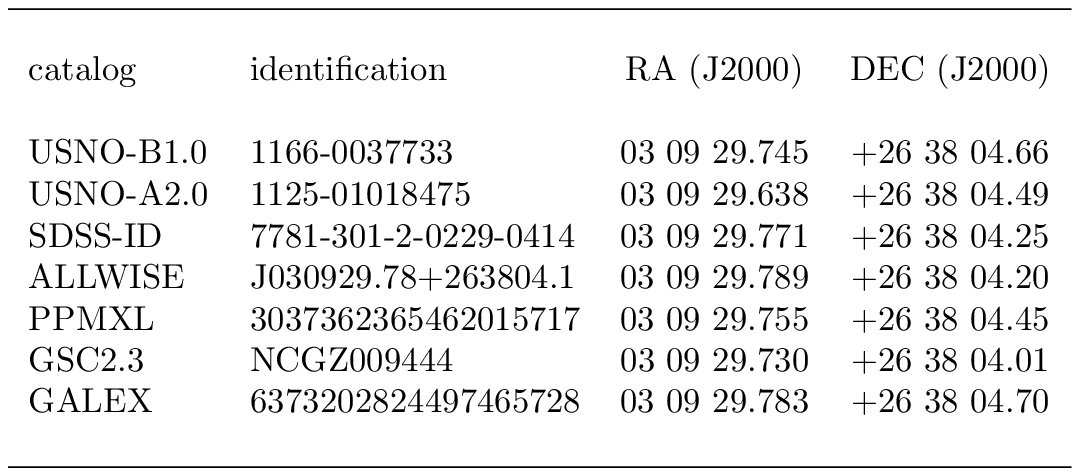}
     \label{tab3}
  \end{table}

  \begin{figure*}[!Ht]
     \centering
     \includegraphics[width=13cm]{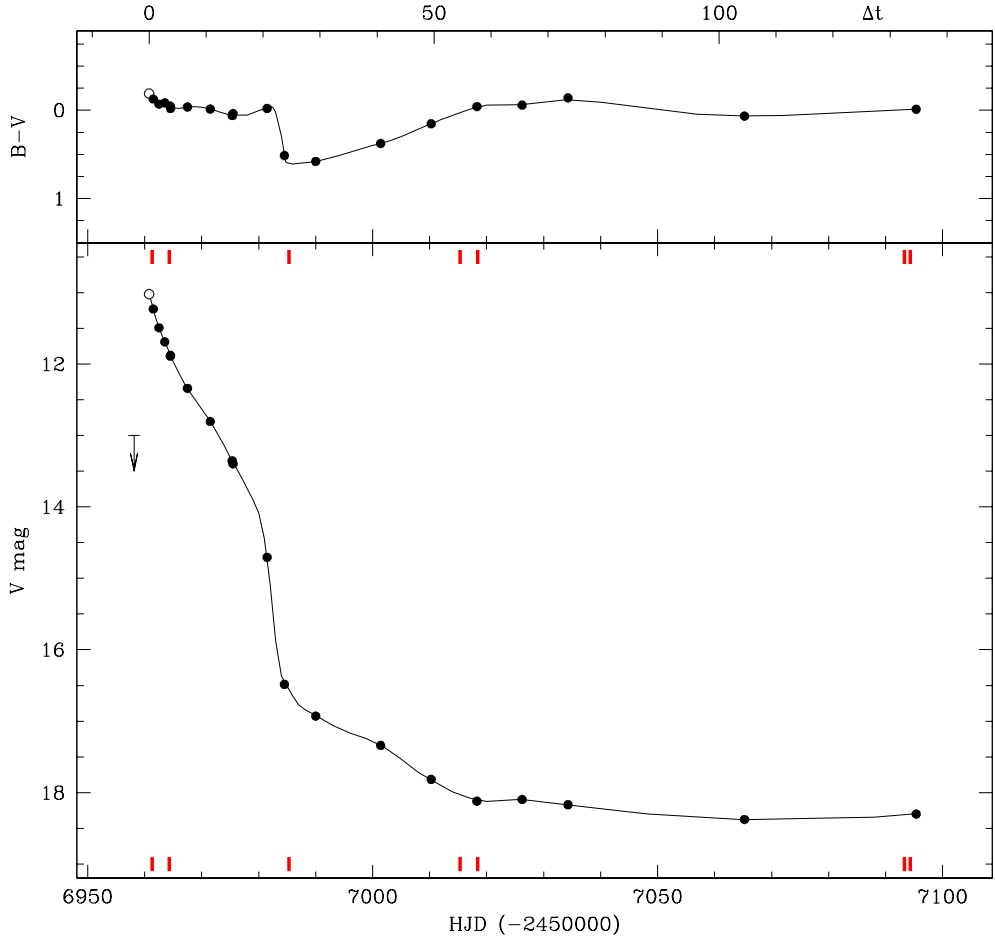}
     \caption{Photometric evolution of PNV J03093063+2638031 during the 2014
     superoutburst, from our observations in Table~1. The thick markers
     point to the epochs of the spectra listed in Table~2. Time on the upper
     abscissae is counted from the epoch of maximum brightness on 2014
     Oct 30.28 UT (JD=2456960.78). For the open circles and the arrow see
     text (sect. 5).}
     \label{fig1}
  \end{figure*}

The photometric observations of PNV during the 2014 superoutburst were
carried out with three different telescopes.  ANS Collaboration telescope N. 
11, located in Trieste (Italy), was used when the object was bright.  It
is a 13cm f/6.6 Vixen ED130SS refractor equipped with Custom Scientific
$U$$B$$V$$R_{\rm C}$$I_{\rm C}$ photometric filters and a SBIG ST10XME CCD
camera, 2184$\times$1472 array, 6.8 $\mu$m pixel providing a focal plane
scale of 1.63$^{\prime\prime}$/pix and a field of view of
60$^\prime$$\times$40$^\prime$.  Technical details and operational
procedures of the ANS Collaboration network of telescopes running since
2005, are presented by Munari et al.  (2012).  Detailed analysis of the
photometric performances and measurements of the actual transmission
profiles for all the photometric filter sets in use at all telescopes is
presented by Munari \& Moretti (2012). During the latest phases of the outburst,
larger instruments were used to measure PNV, namely the Asiago 67/92cm
Schmidt and 1.82m reflector telescopes.  The 67/92cm f/3.2 Schmidt telescope
is equipped with Custom Scientific $B$$V$$R_{\rm C}$$I_{\rm C}$ photometric
filters and a SBIG STL-11000MC2 CCD camera, 4008$\times$2672 array, 9 $\mu$m
pixel providing a focal plane scale of 0.86$^{\prime\prime}$/pix and a field
of view of 58$^\prime$$\times$38$^\prime$.  The 1.82m telescope was used
with the AFOSC spectrograph-imager, that houses $U$$B$$V$$R_{\rm C}$$I_{\rm
C}$ filters and an Andor DW436-BV CCD camera, 2048$\times$2048 array, 13.5
$\mu$m pixel providing a focal plane scale of 0.26$^{\prime\prime}$/pix and
a field of view of 8.9$^\prime$$\times$8.9$^\prime$.  All measurements on
PNV were performed with aperture photometry (the sparse field not requiring
the use of PSF fitting), against a photometric sequence located around the
variable.  The sequence was calibrated from APASS survey data (Henden et al. 
2012, Henden \& Munari 2014) using the transformation equation calibrated in
Munari et al.  (2014).  Our measurements are plotted in Figure~1 and
presented in Table~1, where the reported uncertainties are total error
budgets, combining quadratically the error on the variable with the
uncertainty of the transformation from the istantaneous local photometric
system to the standard one as defined by the local photometric sequence.

Low resolution spectroscopy of PNV was obtained with the 1.22m telescope +
B\&C spectrograph operated in Asiago by the Department of Physics and
Astronomy of the University of Padova.  The CCD camera is a ANDOR iDus
DU440A with a back-illuminated E2V 42-10 sensor, 2048$\times$512 array of
13.5 $\mu$m pixels.  It is highly efficient in the blue down to the atmospheric
cut-off around 3200 \AA, and it is normally not used longward of 8000 \AA\
for the fringing affecting the sensor.  The long-slit spectra were recorded
with a 300 ln/mm grating blazed at 5000 \AA, and extend spatially for 9.5
arcmin (at a scale of 1.119 arcsec/pix) covering the wavelength range from
3300 to 8000 \AA\ at 2.31 \AA/pix.  The spectra were reduced within
IRAF, carefully involving all steps connected with correction for bias, dark
and flat, sky subtraction, wavelength and flux calibration.  The error on
the flux calibration at all epochs should not exceed 6\% anywhere as
the intercalibration of the spectrophotometric standards suggests and the
comparison with the near-simultaneous $B$$V$$R_{\rm C}$$I_{\rm C}$ confirms. 
Table~2 provides a log-book of the spectroscopic observations and the spectra 
of PNV at the first four epochs are plotted in Figure~2.

\section{Progenitor}

At the improved position for PNV derived in sect. 1, a faint star is clearly
visible on both Palomar I and II sky survey images, which was also detected
by various surveys and listed in several catalogs at the positions given in
Table~3.  The straight average of their end figures is 29$^s$.744
($\pm$0.020) and 04$^{\prime\prime}$.39 ($\pm$0.10), coincident within
combined errors with the observed position of PNV.

  \begin{figure*}[!Ht]
     \centering
     \includegraphics[width=15.0cm]{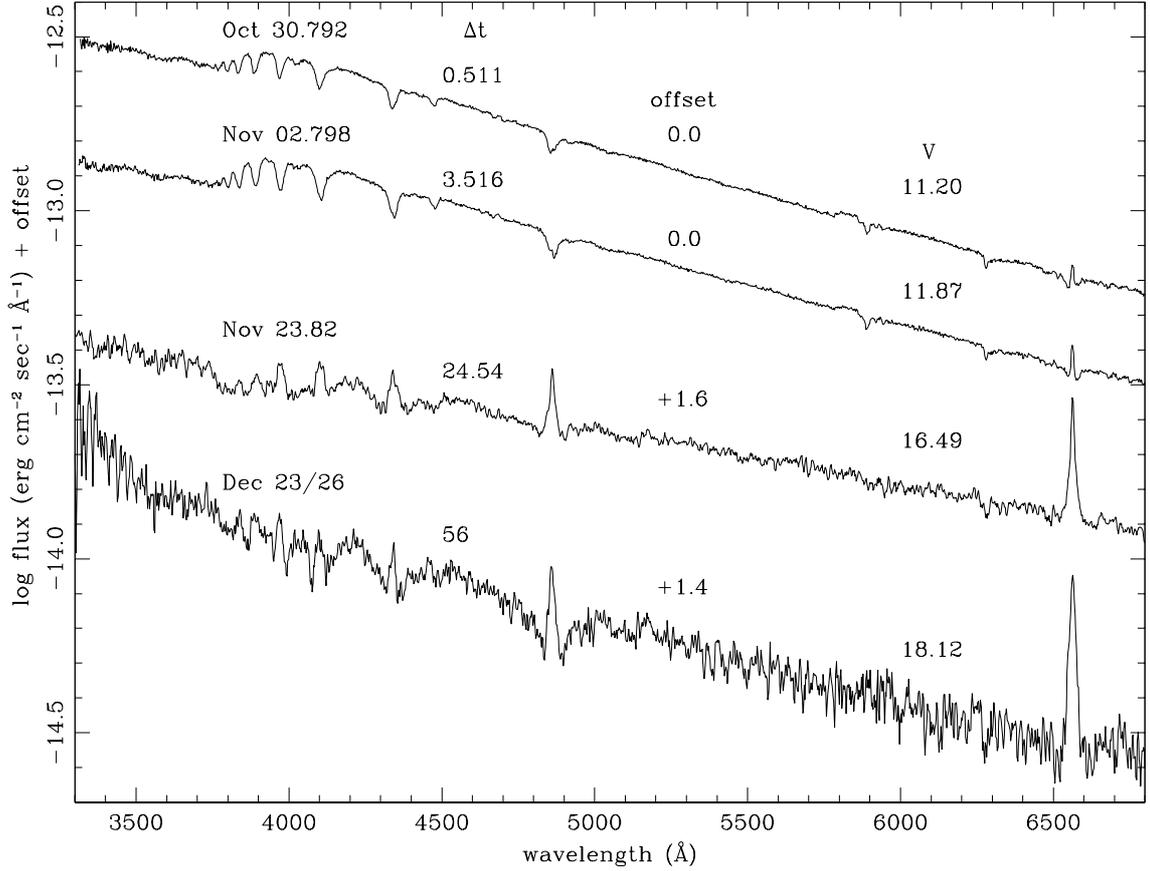}
     \caption{Fluxed spectra of PNV J03093063+2638031 as observed with the
     Asiago 1.22m + B\&C telescope. Note the different slope of Nov 23.824
     spectrum taken at the time of the reddest color displayed by PNV
     J03093063+2638031 (cf Figure~1).}
     \label{fig2}
  \end{figure*}
  \begin{figure}[!Ht]
     \centering
     \includegraphics[width=7.5cm]{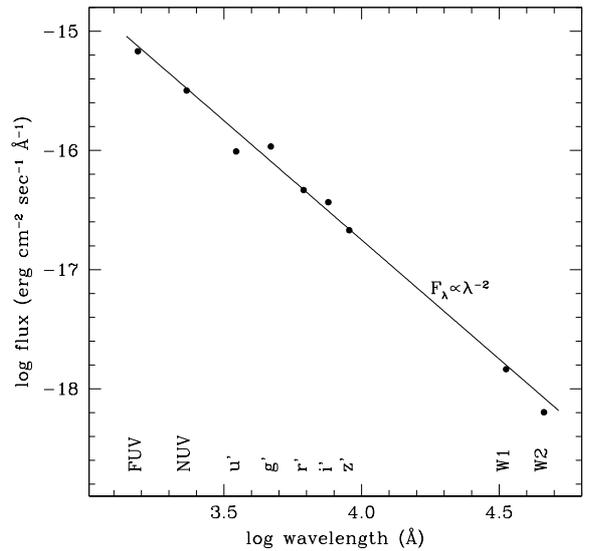}
     \caption{The spectral energy distribution of PNV J03093063+2638031 
     in quiescence, combining non-simultaneous observations by GALEX, SDSS
     and AllWISE.}
     \label{fig3}
  \end{figure}

The PNV progenitor is a faint and blue star. Its USNO-B1.0 magnitudes on
Palomar I and II surveys are $B1$=19.18, $R1$=18.66 and $B2$=19.22,
$R2$=18.14, $I2$=18.70.  The GSC 2.3 catalog reports for PNV $B$=18.80,
$V$=18.25, and $I$=18.35.  The SDSS survey measured it at $u'$=19.137
$\pm$0.031, $g'$=18.882 $\pm$0.009, $r'$=18.907 $\pm$0.012, $i'$=19.011
$\pm$0.017 and $z'$=18.969 $\pm$0.050 mag.  It was also detected by the
GALEX ultraviolet survey at FUV=19.606 $\pm$0.210 and NUV=19.607 $\pm$0.128
mag ($\lambda_{FUV}$=1540 \AA\ and $\lambda_{NUV}$=2315 \AA).  It was too
faint for detection by 2MASS at $J$, $H$ and $K_s$ wavelengths, but it was
detected at mid-infrared wavelengths by AllWISE survey at magnitudes
$W1$=16.397 $\pm$0.082 (S/N=13.3) and $W2$=16.311 $\pm$0.275 (S/N=4.0),
where $\lambda_{W1}$=3.35 $\mu$m and $\lambda_{W2}$=4.6 $\mu$m.  The AllWISE
program (Cutri et al.  2013, VizieR On-line Data Catalog II/328) extends the
work of the WISE space mission (Wright et al.  2010) by combining data from
the cryogenic and post-cryogenic survey phases.

  \begin{figure*}[!Ht]
     \centering
     \includegraphics[angle=270,width=17.0cm]{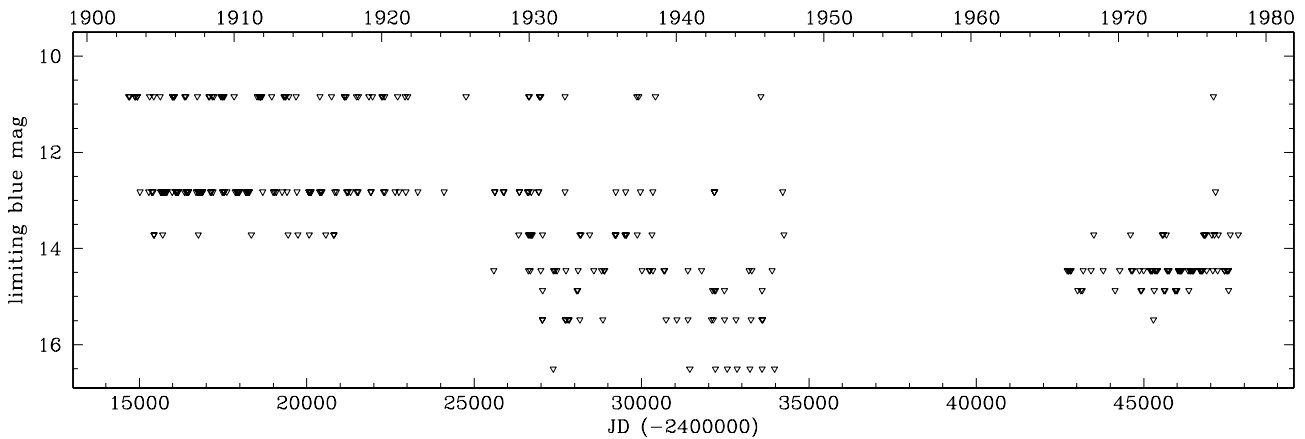}
     \caption{The distribution in time of the {\it "fainter than"} limits
     to the $B$ magnitude of PNV J03093063+2638031 on 332 plates of the
     Harvard plate stack that we inspected looking for previous outbursts.}
     \label{fig4}
  \end{figure*}

The spectral energy distribution of the PNV progenitor is presented in
Figure~4.  It is well fitted by a $F_\lambda \propto \lambda^{-2}$ power law
over the whole UV-opt-IR wavelength range covered by GALEX, SDSS and AllWISE
observations, compatible with an accretion disk dominating the emission from
system in quiescence (La Dous 1989).  The completeness limits of 2MASS
survey in the area around PNV are $J$$\sim$16.5, $H$$\sim$16.0, and
$K$$\sim$15.5 and the corresponding fluxes lie well above the fitting line
in Figure~4, in agreement with the non detection of PNV by 2MASS. For
reference purposes it is useful to list the conventional Johnson-Cousins and
2MASS magnitudes corresponding to the linear fit in Figure~3: $U$=18.76,
$B$=19.51, $V$=19.53, $R_{\rm C}$=19.23, $I_{\rm C}$=19.09, $J$=18.58,
$H$=18.11, and $K_s$=17.55.

\section{Historical lightcurve from Harvard plates}

The superoutbursts in WZ Sge stars are separated by decades, and the normal
outbursts in between are rare or even absent.  To test if this is also the
case for PNV, we went to the Harvard plate stack in Cambridge (USA) to
inspect historical photographs covering the position of PNV.  The vast
majority of the plates at Harvard are blue sensitive and exposed through
lense-astrographs without intervening photometric filters.  In modern terms,
this is roughly equivalent to observe in the $B$ band.  The 2014
superoutburst peaked at $B$$\sim$10.8, so we narrowed our search to plates
going deep enough to reach at least a limiting magnitude of $\sim$10.8 in
blue light.  Out of the $\sim$600 plates that we inspected, 332 of them were
of a photometric quality meeting such criterion, covering the time interval
from Jan 1899 to Nov 1981.  Given the short focal length of most of the
astrographs used to expose the Harvard plates, we augmented the photometric
sequence around PNV (described in sect.2 and optimized for the longer focal
length CCD observations obtained during the 2014 superoutburst) with
additional stars spanning a wider range in magnitude and at greater angular
distances from PNV.  They too were selected from the APASS survey.  The
final sequence extends from $B$=10.152 to 17.155 mag, and it was used at the
microscope eyepiece to estimate the brightness of PNV or the limiting
magnitude of the plate (defined as the mid-point magnitude between the
faintest sequence star still visible and the brightest of those not
visible).

The limiting magnitude of the 332 plates ranged from 10.8 to 16.5. On all of
them PNV was too faint to be detected.  The results are listed in Table~4
(available electronic only) and are plotted in Figure~4.  During the 2014
superoutburst, PNV took less than a month to decline below the limiting
magnitude of the deepest Harvard plates.  Such a fast evolution could easily
accommodate similar events in the time gaps uncovered by the Harvard plates,
including those due to Solar conjunction.  Nonetheless, the very fact that
PNV went undetected on all 332 plates we inspected suggests that its
outbursts are a rare occurrence, in nice agreement with the well established
behavior of WZ Sge stars as a class.

\section{The 2014 outburst}

The photometric evolution of the outburst is presented in Figure~1. It is
closely similar to that seen in the other WZ Sge-type objects (eg.  Nogami
et al.  2009, Chochol et al.  2012).  It is composed of three
distinct parts.

Our observations begun on JD=6961.511, about one day past the original
discovery observations by S.  Ueda, and as soon as darkness allowed after
the announcement of the discovery was posted to TOCP.  We catched the object
already declining from maximum, $\sim$0.2 mag below it.  S.  Kiyota (Japan,
as reported on TOCP) obtained $B$=10.83, $V$=11.02 and $I_{\rm C}$=10.92 on
Oct 30.28 UT (JD=2456960.78), which we take to mark the optical maximum and
that is plotted as an open circle in Figure~1.  According to the discovery
report by S.  Ueda, nothing was visible down to 13 mag at the position of
PNV two days before the discovery (marked by the arrow in Figure~1). 
Therefore, the rise to maximum was (very) fast and, upon hitting maximum
brightness, PNV immediately begun declining.  The initial decline, usually
referred to as the "plateau" in literature, developed at a nearly constant
$B$$-$$V$$\sim$0.0 and took PNV from peak $V$=11.0 down to $V$=14.0 in 19
days.  The spectra of PNV soon after maximum (cf Figure~2) are typical of CV
variables in outburst, with a strong A-type continuum and broad hydrogen
Balmer and HeI absorption lines.  The Balmer continuum is also in absorption
and a weak emission core is visible in H$\alpha$ and marginally in H$\beta$.

This first phase was followed by a very rapid drop in brightness, from
$V$=14.0 to $V$=16.5 in just 4 days, which was accompanied by cooling of the
$B$$-$$V$ color from 0.0 to +0.6 mag.  This change in color is nicely
confirmed by the change in the slope of the continuum of the corresponding
spectrum shown in Figure~2.  Following this rapid drop, the spectrum of PNV
developed strong hydrogen Balmer emission lines, the absorption in the
Balmer continuum vanished and early hints emerged of the very wide hydrogen
Balmer absorption lines from the underlying white dwarf.  The apparent
reinforcing of the Balmer emission lines was actually driven by the rapid
drop in intensity of the underlying continuum.  In fact, the integrated flux
of the H$\alpha$ emission line for the four spectra presented in Figure~2
declined steadily with time: it was 105, 81, 9.8 and 5.9$\times$10$^{-15}$
erg cm$^{-2}$ sec$^{-1}$, respectively.

The third and last phase saw PNV to slowly complete the final decline toward
quiescence brightness, in pace with a gradual warming of the $B$$-$$V$ color
from +0.6 back to $\sim$0.0 mag.  This phase took PNV from $V$=16.5 to
$V$=18.1 in $\sim$35 days, and to $V$=18.4 in additional $\sim$45 days.
This is still one mag brighter than the quiescence level, taken to be
represented by the linear fit of Figure~3 (see sect. 3), suggesting that the
return to proper quiescence required significant additional time. Our
spectrum of PNV obtained when its was $V$=18.1 mag, displays strong and very
wide Balmer absorption lines from the underlying white dwarf. H$\alpha$
emission completely fills in the corresponding absorption line, while higher
Balmer lines are reduced to just emission cores. This is the combined effect
of a reduction in the integrated flux of emission lines for higher term
Balmer lines (5.9, 3.7, and 1.9$\times$10$^{-15}$ erg cm$^{-2}$ sec$^{-1}$
for H$\alpha$, H$\beta$ and H$\gamma$, respectively) and the underlying
white dwarf spectrum rapidly increasing in intensity toward the blue.  No
terminal {\em echo} outbursts, of the type displayed for example by EG Cnc
(Patterson et al. 1998), was observed in PNV during the return to quiescence
brightness (cf Figure~1).

  \begin{table}[!Ht]
     \centering
     \caption{Upper limits to the magnitude of PNV J03093063+2638031 on the
     332 Harvard plates we inspected looking for previous unrecorded
     outbursts. The table is published in its entirety in the electronic
     edition of this journal. A portion is shown here for guidance regarding
     its form and content.}
     \includegraphics[width=6.8cm]{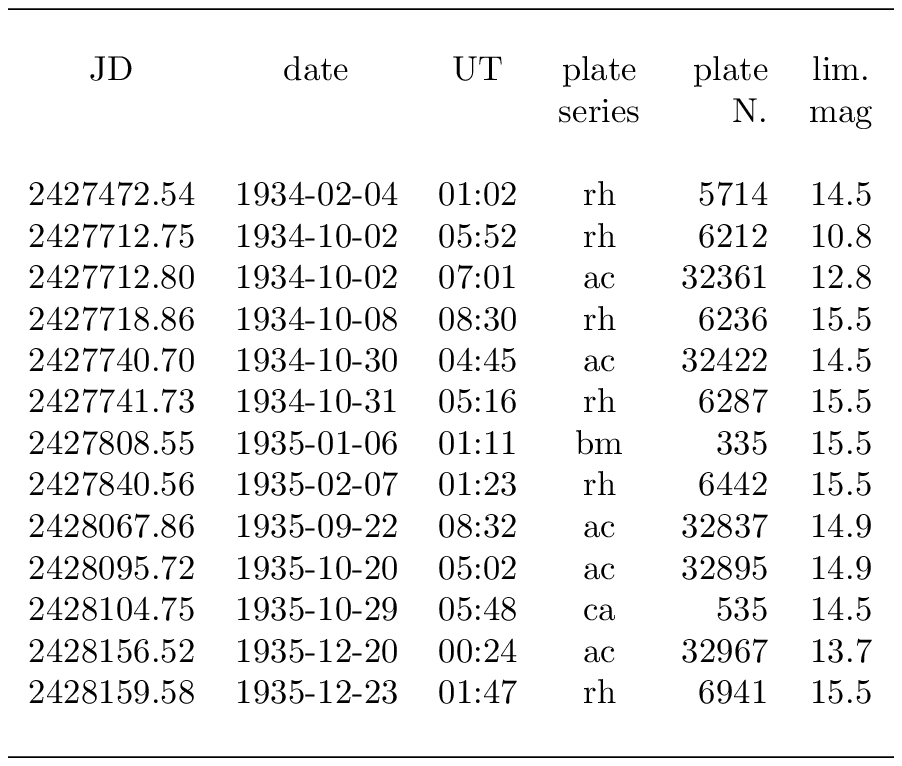}
     \label{tab4}
  \end{table}

  \begin{figure*}
     \centering
     \includegraphics[width=16cm,angle=270]{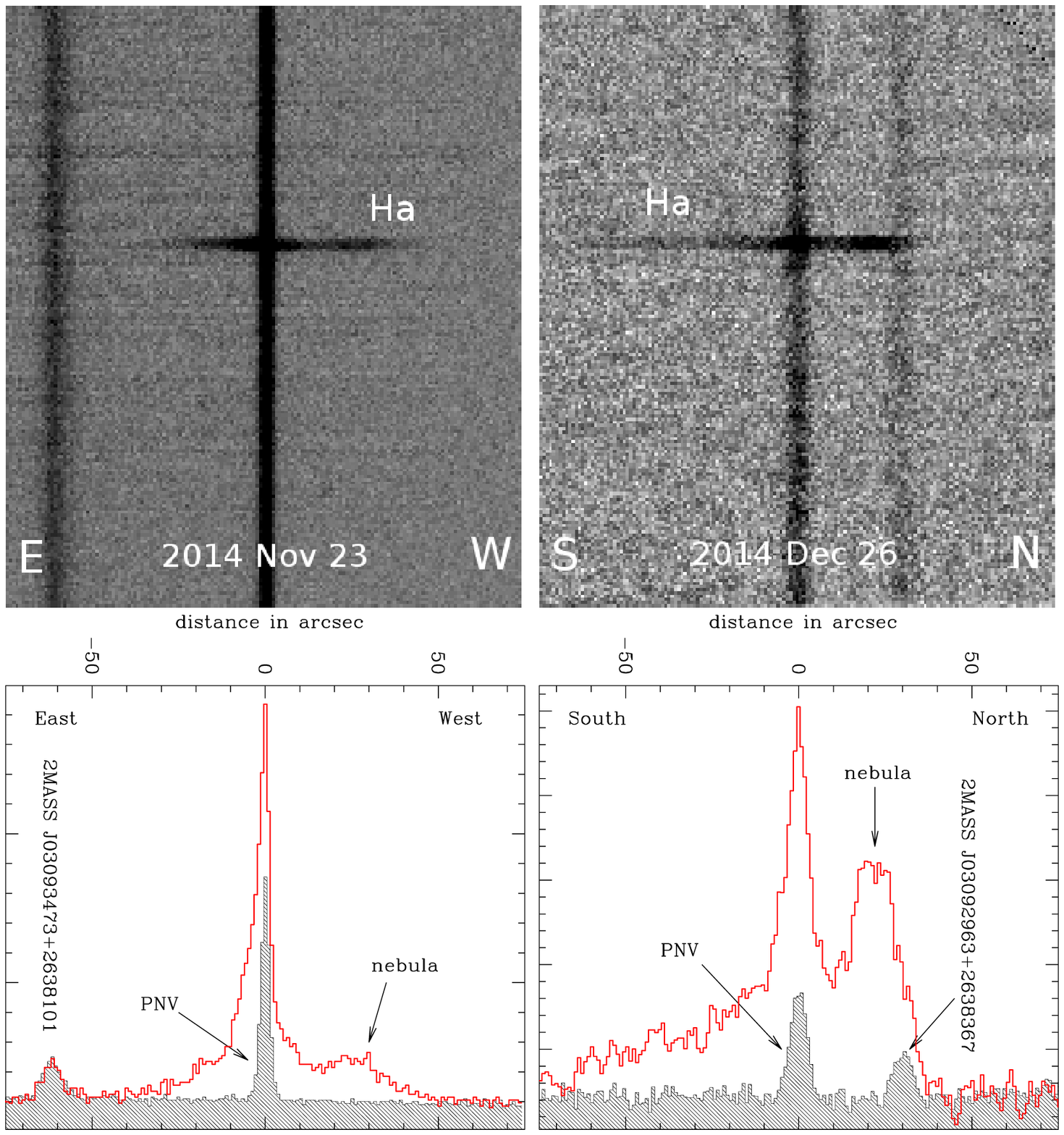}
     \caption{{\em Top left:} part of the 2D low-resolution spectrum of PNV
     J03093063+2638031 we obtained on 23 Nov 2014, when the object was
     shining at $V$=16.7 during the advanced decline from outburst maximum. 
     The wavelength range covers from 6393 \AA\ at the top to 6821 \AA\ at
     the bottom and shows the H$\alpha$ spatially extended nebula as
     recorded with the spectrograph slit alligned East-West.  The weak and
     fuzzy spectrum at left is that of 2MASS 03093473+2638101, a nearby
     $R$=12.4 mag star barely intercepted by the slit.  {\em Bottom left:}
     tracing along the spatial axis of the same pectrum of PNV
     J03093063+2638031 shown on the top panel.  The thicker red line traces
     along the emission from the nebula, while the shaded profile traces
     just away from the nebula to provide a reference.  {\em Right panels:}
     the same for the 26 Dec 2014 spectrum, this time with the slit oriented
     North-South and when PNV J03093063+2638031 was at $V$=18.1 mag. The
     weak spectrum at right is that of 2MASS 03092963+2638367, a nearby
     $R$=17.5 mag star marginally intercepted by the slit.}
     \label{fig5}
  \end{figure*}

\section{Discovery of a large, spatially resolved and outburst-ionized
emission nebula around PNV J03093063+2638031}

A very peculiar, probably unique feature of PNV among known WZ Sge-type
stars, is the detection of a spatially extended nebula centered on the
variable that became briefly visible because ionized by the 2014 outburst,
proving its firm physical association to PNV.  We discovered it during the
spectroscopic observations with the Asiago 1.22m telescope, carried out in
long-slit mode with the slit height extending for about 9.5 arcmin on the
sky.  The nebula is not of the reflection type (thus producing a {\it
light-echo} of the outburst), because it is visible only in the Balmer
emission lines and not in the continuum (which was brighter than the
emission lines around outburst maximum).

On the first observation on 30 Oct 2014 (day +0.51) there is no hint of
spectral emission features spatially extending more than the stellar
continuum.  However, on the next spectrum obtained on 2 Nov, day +3.52, the
H$\alpha$ emission was already noticeably extended, in a similar manner on
both sides of the stellar spectrm.  The spatial FWHM of the stellar spectrum
(dominated by seeing dispersion) was 3.92 pixels, that of the H$\alpha$ was
11.62 pixels.  This corresponds to a seeing-corrected FWHM(H$\alpha$)=12.2
arcsec.

From this we may estimate a distance to PNV if we assume that this extended
H$\alpha$ emission is produced within a spherical and homogeneous nebula
centered on PNV, that begins to emit by recombination as soon as the
ionizing radiation from the outbursting central star reaches it.  The time
passed from discovery (when the object was already close to maximum
brightness) to our spectroscopic detection of the extended nebula was 4.17
days.  Assuming that the rise to maximum of PNV was nearly istantaneous,
this means that the outer radius of the ionized volume was at that
time 1.1$\times$10$^{16}$ cm from PNV, or 3.5$\times$10$^{-3}$ pc. Comparing
with a FWHM of 12.2 arcsec, the distance turns out to be $\sim$120 pc, and
the absolute magnitude in quiescence is M$_V$=14.1 (assuming a negligible
extinction at the high galactic latitude of PNV).

Next time we observed again PNV it was 21 days later (Nov 23), when the star
was $\sim$4.6 mag fainter. The bright {\it plateau} past maximum was over
by $\sim$6 days and with it the sustained input of ionizing photons to the
circumstellar medium. By this time the spatial extent of the ionized
nebula had significally grown. The appearance of the spatially extended
H$\alpha$ emission on this spectrum is shown on the left panel of Figure~4.
The outer portion of the nebula extends to 40 arcsec toward East and 50
arcsec to the west of PNV, with considerable substructures visible in the
intensity profile, indicating large departures from spherical symmetry
in the distribution of the gas.

A month later, we invested considerable observing time at two distinct and
nearby dates (Dec 23 and 26) to obtain long-slit spectra of PNV at two
perpendicoular orientations, East-West and North-South.  The spatially
extended H$\alpha$ emission along the East-West was similar to that of a
month earlier, only with a reduced brightness.  This indicates that not much
gas exists external to that already ionized by Nov 23 (density bounded
nebula), and that the ionized gas was already completing the recombination
given the long gone photo-ionization from the central star.  The spatially
extended profile along the North-South direction for Dec 26 observations is
shown on the right panel of Figure~4.  The H$\alpha$ profile along the
North-South directions extends by similar amounts compared to East-West: 35
arcsec toward North and 65 arcsec toward South, indicating that PNV sits
approximately at the projected center of the nebula.  Considerable
sub-structures are visible also along the North-South spatial extention of
the nebula, as it is for the East-West.

To check on the photo-ionization/recombinations scenario for the nebula, we
re-observed PNV with the same instrumentation in March 2015, exposing much
longer than previous visits. No trace of the nebula was more visible,
indicating that recombination within the ionized gas was by that time
completed and the circumstellar nebula had returned to obscurity from which
it had briefly emerged during the superoutburst of 2014. The great surface
brightness it displayed (expecially at the time we discovered it) would have
resulted in an easy detection on the red Palomar Sky Survey plates, but they
shows just empty sky at the position of the nebula.

\section{Discussion}

At the limited spectral resolution of our spectra, the spectral width of the
spatially extended H$\alpha$ is comparable to that of the lines from the
NeArFe calibration lamp or the night-sky lines (their FWHM corresponds to
$\sim$250 km/sec at H$\alpha$ wavelengths).  This means that the local
velocity dispersion within the circumstellar gas is low, in particoular the
gas is not fast expanding.  The nebula also does not rotate with a large
angular velocity, nor it is subject to large scale motions because its
emission is centered at the same wavelength along its whole spatial extent. 
In short, the circumstellar nebula looks like a quiet blob of gas, with
considerable sub-structures and a clear physical association to PNV.

The hydrogen recombination time scale (in hours) is related to electron
temperature and density by:
\begin{equation}
t_{\rm rec} = 0.66 \left(\frac{T_{\rm e}}{10^4 {\rm ~K}}\right)^{0.8}
\left(\frac{n_{\rm e}}{10^9 {\rm ~cm}^{-3}}\right)^{-1}
\end{equation}
following Ferland (2003). With an observed recombination time scale of the
order of $\sim$two months, the electron density is 2.6, 4.5 or
7.9$\times$10$^5$ cm$^{-3}$ for $T_e$=5000, 10000 or 20000 K, respectively.

With a mean angular extension of 50 arcsec and a distance of 120 pc, the
physical radius of the nebula is 0.03 pc, or 1/10 of a typical planetary
nebula.  If homogeneously filled by hydrogen at the density estimated above
for $T_e$=10000 K, its mass would be 1.1 M$_\odot$.  The marked
sub-structures visible in Figure~4 suggest that a sizeable fraction of the
available volume is indeed empty.  Even if the filling factor would be just
a tiny 1\%, still the amount of mass in the nebula would be orders of
magnitude greater than that ejected by outbursts on white dwarfs, like nova
eruptions that - depending on the mass of the white dwarf - are observed to
expel from 10$^{-6}$ to 10$^{-4}$ M$_\odot$ (Bode \& Evans 2008). 
Furthermore, the ejection velocities observed in novae (from several
hundreds to a few thousand km/s, Warner 1995) are much larger than the low
(if any) expansion velocity observed for the nebula around PNV.

Appreciable mass ejection cannot be expected from WZ Sge-type binaries,
based on their nature.  Such systems are composed by a white dwarf orbited
every $\sim$80 min by a brown dwarf, the mass of the latter being of the
order of a mere $\sim$0.06 M$_\odot$ (Hellier 2001), below the limit for
stable hydrogen burning in the core.  The mass transfer rate in such system
is extremely low, $\sim$10$^{-12}$ M$_\odot$/yr, and the mass reservoir in
the donor star virtually null.  Thus, it is highly improbable that the
nebula around PNV originated from the brown dwarf companion.

It could be speculated that the nebula around PNV is the remnant of the slow
wind the AGB progenitor blown off before turning into the present day white
dwarf.  There are however several problems with this origin too.  The
dilution of AGB wind into the surrounding instertellar space is rapid, of
the order of 10$^4$ yrs which is the age of common planetary nebulae, and
the evolutionary path leading to the formation of WZ Sge-type binaries is
instead much longer than that ($\sim$10$^9$ yrs).  Furthermore, AGB winds
produce a lot of dust.  In fact, planetary nebulae show clear signs of
thermal emission from dust mixed with the gas, but the 2MASS and AllWISE
data indicate that no detectable dust is present in PNV.

The circumstellar nebula that should not exist is instead stubbornly there,
with a clear physical association to PNV as proven by the sudden
photo-ionization induced by the 2014 superoutburst.  In light of the
challenges this nebula seems to pose to theoretical models, it is highly
advisable to search for similar nebulae around future WZ Sge-type
transients.  This will require spectrographs equipped with long slits and
the investment of long integration times, not easy to accommodate in present
times dominated by remote service observing and multi-object, fiber-fed
spectrographs.

\begin{acknowledgements}
We would like to thank Alison Doane, Curator of Astronomical Photographs at
the Harvard College Observatory, for granting access to the Harvard plate
stack, and Stella Kafka for hospitality at AAVSO Headquarters.  This work
was supported in part by the University of Rijeka under the project number
13.12.1.3.03.  and by the Croatian Science Foundation under the project 6212
Solar and Stellar Variability.
\end{acknowledgements}

\end{document}